\documentclass[a4paper,11pt]{article}
\usepackage{pos}

\usepackage[italian,english]{babel}
\usepackage[utf8]{inputenc}
\usepackage[T1]{fontenc}

\usepackage{xcolor}

\newcommand{\Nf}{{N_f}}
\newcommand{\psibar}{{\bar{\psi}}}
\newcommand{\corr}[1]{\langle #1\rangle}
\newcommand{\mathsp}{\;\;\;\;}

\newcommand{\bs}[1]{\boldsymbol{#1}}
\newcommand{\csw}{c_{\text{sw}}}
\newcommand{\LambdaQCD}{\Lambda_{\text{QCD}}}

\newcommand{\Zgs}[1][6]{Z_G^{\{#1\}}}

\newcommand{\Zfs}[1][6]{Z_F^{\{#1\}}}

\newcommand{\Tgs}[1][\mu\nu]{T^{G,\{6\}}_{#1}} 
\newcommand{\Tgt}[1][\mu\nu]{T^{G,\{3\}}_{#1}}
\newcommand{\Tfs}[1][\mu\nu]{T^{F,\{6\}}_{#1}}
\newcommand{\Tft}[1][\mu\nu]{T^{F,\{3\}}_{#1}}

\newcommand{\vxi}{\boldsymbol{\xi}}

\newcommand{\vxithzero}{ {\scriptscriptstyle \vxi,\theta_{\scalebox{0.50}{0}}} }

\usepackage{amsfonts} 
\usepackage{amsmath}
\usepackage{mathtools}

\usepackage{graphicx}

\usepackage{subcaption}

\usepackage{hyperref}

\usepackage{multirow}

\title{Thermal QCD for non-perturbative renormalization of composite operators}
\ShortTitle{Thermal QCD for non-perturbative renormalization of composite operators}

\author*[a,b]{Matteo Bresciani}
\author[c]{Mattia Dalla Brida}
\author[a,b]{Leonardo Giusti}
\author[b]{Michele Pepe}

\affiliation[a]{Dipartimento di Fisica, Università di Milano-Bicocca,\\
Piazza della Scienza 3, I-20126 Milano, Italy}

\affiliation[b]{INFN, Sezione di Milano-Bicocca,\\
Piazza della Scienza 3, I-20126 Milano, Italy}

\affiliation[c]{Theoretical Physics Department, CERN\\
  1211 Geneva 23, Switzerland}

\emailAdd{Matteo.Bresciani@mib.infn.it}
\emailAdd{mattia.dalla.brida@cern.ch}
\emailAdd{Leonardo.Giusti@mib.infn.it}
\emailAdd{Michele.Pepe@mib.infn.it}

\abstract{
We present our progresses in the use of the non-perturbative renormalization framework based on considering
QCD at finite temperature with shifted and twisted (for quarks only) boundary conditions in the compact direction. We report our final results in the application of this method for the non-perturbative renormalization of the flavor-singlet local vector current. We then discuss the more challenging case of the renormalization of the energy-momentum tensor, and show preliminary results on the relevant one-point functions for the computation of the renormalization constants of its non-singlet components.

\begin{flushright}
CERN-TH-2022-184
\end{flushright}
}

\FullConference{%
The 39th International Symposium on Lattice Field Theory,\\
8th-13th August, 2022,\\
Rheinische Friedrich-Wilhelms-Universität Bonn, Bonn, Germany
}

\begin{document}
\maketitle

\numberwithin{equation}{section}

\graphicspath{{./images/}}
	
\section{Introduction}
Quantum chromodynamics (QCD) is the field theory describing the strong interactions among quarks and gluons. The lattice formulation of this theory is the only known regularization where the properties of QCD can be studied non-perturbatively from first principles. In this setup the inverse of the lattice spacing $a$ provides the ultraviolet cutoff of the theory, and it is removed by taking the continuum limit $a\to0$ of lattice-defined quantities. In general, these quantities have to be properly renormalized in order to recover the correct continuum limit. For example, if we consider a conserved quantity related to a symmetry of continuum QCD, its discretized counterpart may not be conserved because the lattice may break that symmetry and recover it in the continuum limit. Given a lattice operator $O$, we define its renormalized counterpart as
\begin{equation}
	O^R = Z_O \left[ O + \sum_k \frac{c_k}{a^{d-d_{k}}} O_k \right]\;, \mathsp d=\text{dim}(O)
	\label{eq:OR}
\end{equation}
which is a linear combination of the original (bare) operator and eventually of some other lattice operators with the same symmetries and lower or equal mass dimension.
The operator $O^R$ is determined non-perturbatively once the coefficients appearing in its definition \eqref{eq:OR}, which we call renormalization constants, are computed non-perturbatively on the lattice.

Several different renormalization schemes have been proposed to accomplish this task: the Schrödinger Functional scheme \cite{Luscher:1992an}, the RI-MOM scheme \cite{Martinelli:1994ty}, and the Wilson flow scheme \cite{Luscher:2010iy}. We present here the results of the application of a non-perturbative renormalization scheme based on considering the Euclidean formulation of QCD at finite temperature with shifted boundary conditions for the link field, and shifted and twisted boundary conditions for the quark fields. This setup was first proposed in \cite{Giusti:2010bb,DallaBrida:2020gux}, where it was shown that the partition function of a relativistic thermal field theory can be represented as a Euclidean path integral with shifted and twisted boundary conditions on the fields. In the particular case of lattice QCD at finite temperature these boundary conditions are \cite{DallaBrida:2020gux}
\begin{equation*}
	U_\mu(x_0+L_0,\bs{x}) = U_\mu(x_0,\bs{x}-L_0\bs{\xi})
\end{equation*}
\begin{equation}
	\psi(x_0+L_0,\bs{x}) = -e^{i \theta_0}\, \psi(x_0,\bs{x}-L_0\bs{\xi})
	\label{eq:shtwBCs}
\end{equation}
\begin{equation*}
	\psibar(x_0+L_0,\bs{x}) =-e^{-i \theta_0}\, \psibar(x_0,\bs{x}-L_0\bs{\xi})
\end{equation*}
where $L_0$ is the temporal extension of the lattice, $\vxi$ is the shift vector and $\theta_0$ is the twist phase. They are equivalent to consider the thermal theory in a moving frame with Euclidean boost $\vxi$, at the temperature $T=\gamma/L_0$ where $\gamma=1/\sqrt{1+\vxi^2}$, and with an imaginary chemical potential $\mu_{\cal I}=-\theta_0/L_0$. In \cite{Giusti:2011kt,Giusti:2012yj,DallaBrida:2020gux} this description was further explored with the derivation of some Ward Identities which are non-trivial in presence of shifted boundary conditions and which allow the determination of many thermodynamic properties of the system. In those papers it was also observed that these Ward Identities could be used to determine in a non-perturbative way the renormalization constants of the energy-momentum (EM) tensor of lattice QCD.

This renormalization scheme was successfully employed for the non-perturbative renormalization of the EM tensor in the SU$(3)$ Yang-Mills theory \cite{Giusti:2015daa}, and for the non-perturbative determination of the Equation of State of the same theory \cite{Giusti:2016iqr}. We have applied this new strategy for the first time in full QCD for the non-perturbative renormalization of the flavor-singlet local vector current \cite{Bresciani:2022lqc}, as presented in Section \ref{sec:ZV}. This is a new result since no non-perturbative determination of this renormalization constant has been carried out so far with Wilson fermions (some results with staggered quarks can be found in \cite{Hatton:2020vzp}). Moreover, the renormalization of the flavor-singlet local vector current allowed us to experiment in QCD our renormalization scheme in a simple but non-trivial exercise, without worrying about all the technicalities involved in the renormalization of the QCD energy-momentum tensor. We are currently working on the latter by exploiting all the properties of thermal QCD with shifted and twisted boundary conditions. In Section \ref{sec:Z_Tmunu} we describe our strategy, whose effectiveness is confirmed by the numerical values we are obtaining from the lattice calculations.

\section{Renormalization of the flavor-singlet local vector current}
\label{sec:ZV}
In the continuum, the flavor-singlet vector current
\begin{equation}
	V_\mu(x) = \psibar(x)\gamma_\mu\psi(x)
	\label{eq:V}
\end{equation}
is a conserved quantity related to the invariance of QCD under the vector subgroup U$(1)_V$ of the chiral symmetry. This subgroup is respected by the lattice regularization too, and the flavor-singlet conserved vector current can be derived through Noether's theorem applied to the discretized QCD action. In case of Wilson fermions its expression is the following:
\begin{equation}
	V^c_\mu(x)=	\frac{1}{2}\left[\bar{\psi}(x+a\hat{\mu})U^\dagger_\mu(x)\left(\gamma_\mu+1\right)\psi(x)
				+\bar{\psi}(x)U_\mu(x)\left(\gamma_\mu-1\right)\psi(x+a\hat{\mu})\right]
	\label{eq:Vc}
\end{equation}
Despite the exact continuum behavior of $V^c_\mu$, the computation of this current requires the evaluation of the quark fields in two neighboring points of the lattice, which may result in noisy measurements of this lattice operator and larger discretization effects. It is therefore worthwhile to consider the so-called flavor-singlet local vector current $V^l_\mu$, obtained by the naive discretization of the continuum current \eqref{eq:V}. This lattice current has no point-split problem, but it must be renormalized. We can determine its renormalization constant $Z_V$ by comparing the one-point functions of the conserved and of the local currents:
\begin{equation}
	Z_V(g_0^2) = \lim_{a/L_0\to0}\left.\frac{\corr{V^c_\mu}_\vxithzero}
			               {\corr{V^l_\mu}_\vxithzero}\right|_{g_0^2,L_0/a}
	\label{eq:ZVnoimpr}
\end{equation}

\begin{figure}
	\centering
	\begin{subfigure}{.5\textwidth}
		\centering
		\includegraphics[scale=0.45, trim=0cm 0.9cm 0cm 0cm clip]{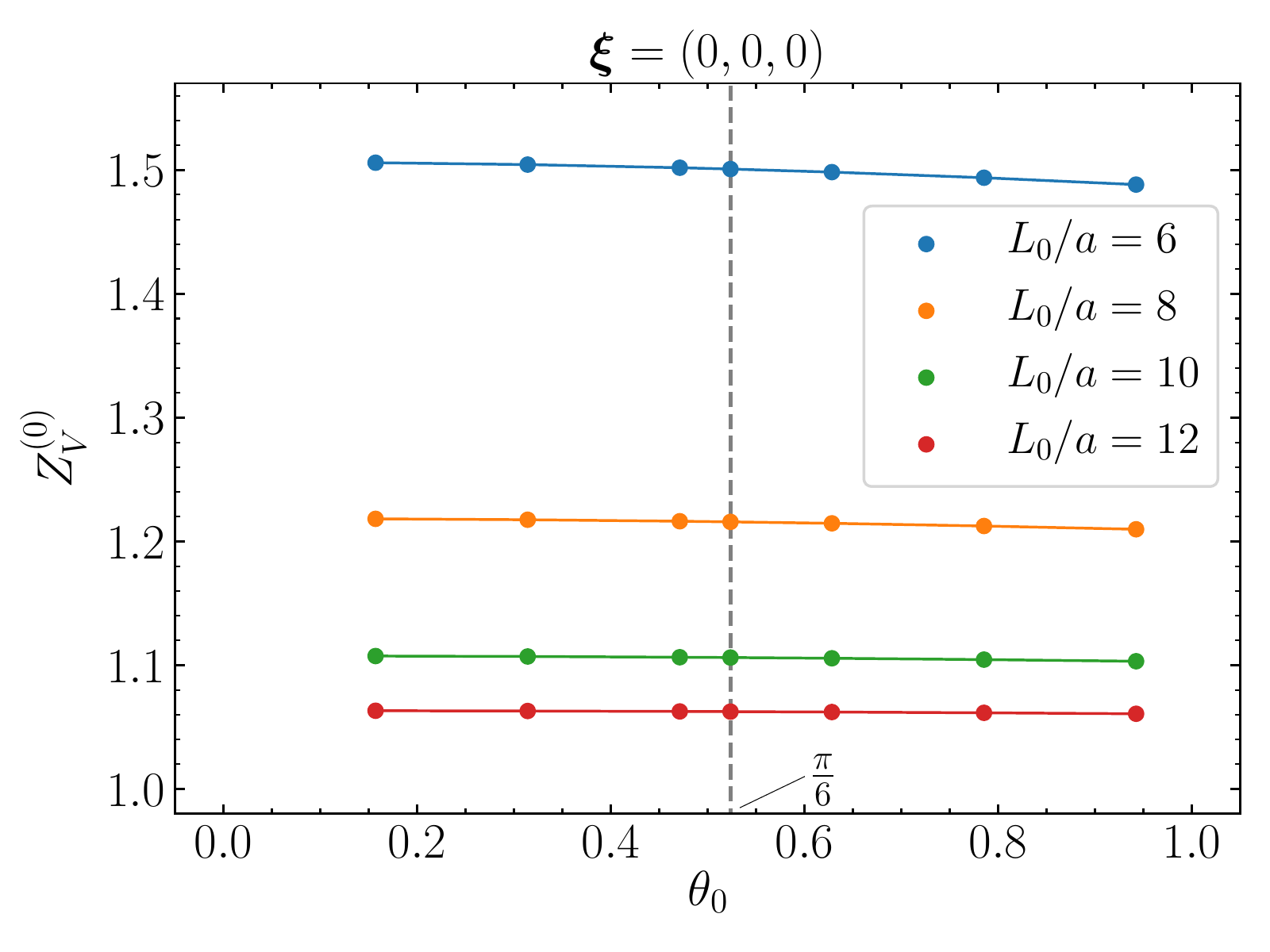}
	\end{subfigure}%
	\begin{subfigure}{.5\textwidth}
		\centering
		\includegraphics[scale=0.45, trim=0cm 0.9cm 0cm 0cm clip]{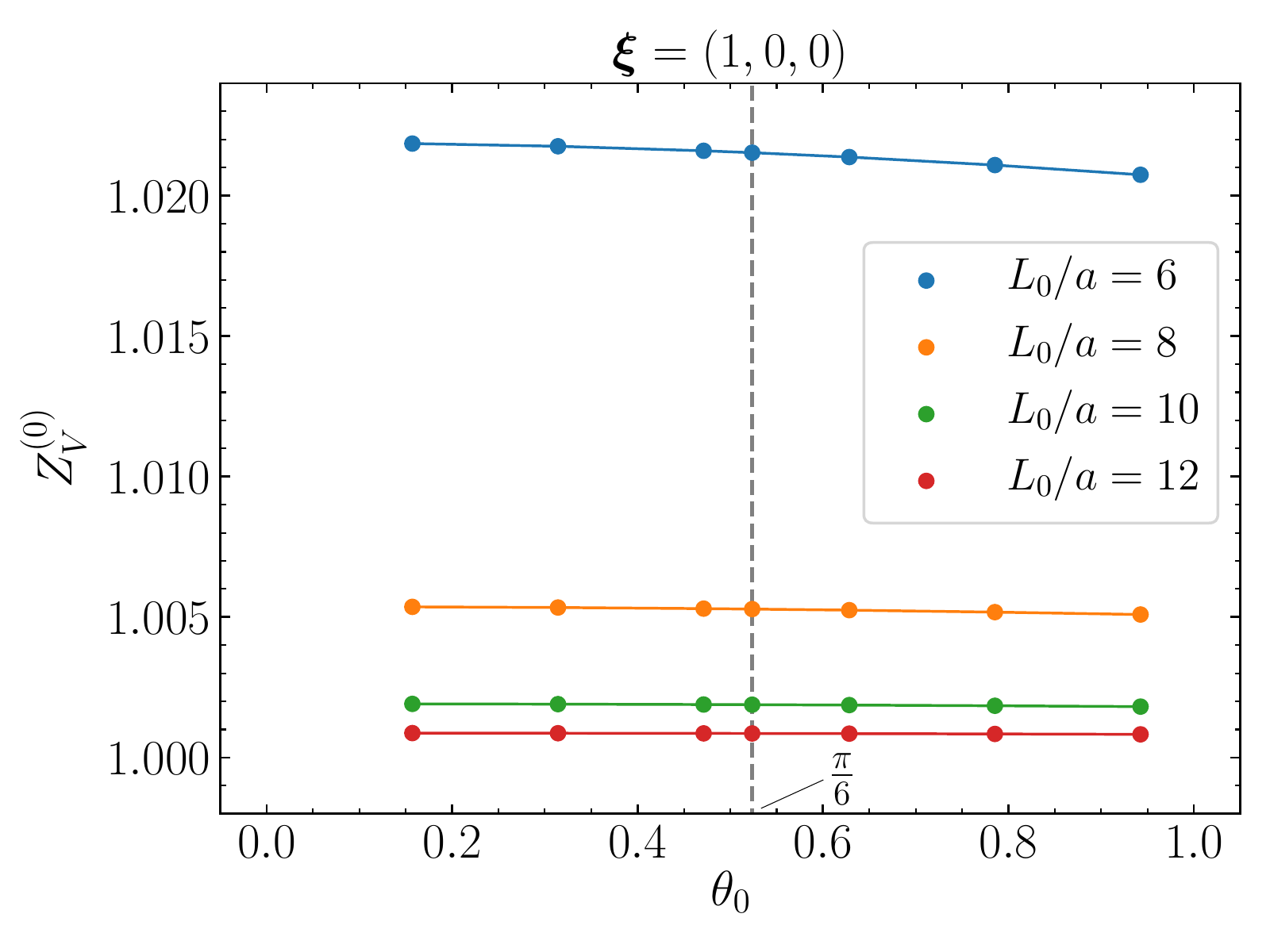}
	\end{subfigure}
	\caption{Tree-level $Z_V$ as a function of the twist phase $\theta_0$, at zero shift and $\vxi=(1,0,0)$.}
	\label{fig:ZVart}
\end{figure}

The ratio on the right is taken at given bare coupling $g_0$, so at given lattice spacing $a$, and in the thermodynamic limit. We introduce a shift of value $\vxi=(1,0,0)$: this is not strictly required but it turns out that, compared to periodic boundary conditions, discretization effects are much reduced, as we can see from the tree-level values of $Z_V$ plotted in Figure \ref{fig:ZVart}. The same figure shows that the dependence on the twist phase $\theta_0$ is very mild. The Euclidean QCD partition function with an imaginary chemical potential enjoys an effective $2\pi/3$ periodicity in $\theta_0$ \cite{Roberge:1986mm}, and we choose $\theta_0=\pi/6$ which is in the middle of the range $[0,\pi/3]$. The presence of the twist phase is mandatory because otherwise the expectation values of the vector currents appearing in \eqref{eq:ZVnoimpr} would be zero.

The use of the O$(a)$-improved Wilson action, together with the O$(a)$-improved fields \cite{Bhattacharya:2005rb}
\begin{equation}
  	\hat{V}_\mu^{c,l} (x) = V_\mu^{c,l} (x) - \frac{a}{4}\, c_V^{c,l}\,
  	(\partial_\nu+\partial^*_\nu) \left(\psibar(x)\, [\gamma_\mu,\gamma_\nu] \, \psi(x) \right)
 	\label{eq:Oa_impr}
\end{equation}
guarantees that the the matrix elements $\corr{V_\mu^{c,l}}$ are O$(a)$-improved. The coefficients $c_V^{c,l}$ can be determined so that O($a$) lattice artifacts are removed from some convenient $n$-point function of the improved fields \eqref{eq:Oa_impr}; then any correlator of these fields will be improved because the $c_V^{c,l}$ are independent on the particular correlator.
Actually, since we are interested in one-point functions of the currents, the contributions from the improving terms vanish for translation invariance and therefore the $\corr{V_\mu^{c,l}}$ are automatically O($a$)-improved.

As a technical tool to make the extrapolations milder, we can further improve our definition \eqref{eq:ZVnoimpr} for $Z_V$ by subtracting the difference between the 1-loop perturbative $Z_V$ in the $a/L_0\to0$ limit with the same quantity at fixed temporal extension. At given $a/L_0$ we get
\begin{equation}
	Z_V\left(g_0^2, \frac{a}{L_0}\right) = \left.\frac{\corr{V^c_\mu}_\vxithzero}
			               {\corr{V^l_\mu}_\vxithzero}\right|_{g_0^2,L_0/a}
				+\left[1 + c_1g_0^2-Z_V^{(0)}\left(\frac{a}{L_0}\right)
				\left(1+g_0^2\,\frac{8}{3}Z_V^{(1)}\left(\frac{a}{L_0}\right)\right)\right]
	\label{eq:ZV}
\end{equation}
The 1-loop coefficient $c_1$ is \cite{Skouroupathis:2008mf}
\begin{equation}
	c_1 = \frac{1}{12\pi^2}\left[-20.617798655(6) + 4.745564682(3)\,\csw + 0.543168028(5)\,\csw^2\right]
	\label{eq:1loop_coeff}
\end{equation}
and, at the relevant order, $\csw=1+0.26590(7)g_0^2$ \cite{Luscher:1996vw}. The values of $Z_V^{(0)}$, $Z_V^{(1)}$ in 1-loop lattice perturbation theory can be found in Table \ref{tab:ZVpert}.
\begin{table}
	\centering
	\begin{tabular}{ccc}
	\hline
	$L_0/a$ & $Z_V^{(0)}$ & $Z_V^{(1)}$ \\
	\hline
	 4 & 1.112904 & -0.057954 \\
	 6 & 1.021530 & -0.051313 \\
	 8 & 1.005285 & -0.049255 \\
	10 & 1.001882 & -0.048787 \\
	\hline
	\end{tabular}
	\caption{Perturbative values of $Z_V$ at given $L_0/a$, in the thermodynamic limit, at $\theta_0=\pi/6$ and
	         $\vxi=(1,0,0)$.}
	\label{tab:ZVpert}
\end{table}

In the $a/L_0\to0$ limit we can parameterize the dependence of $Z_V$ on the lattice artifacts as follows:
\begin{equation}
	Z_V\left(g_0^2, \frac{a}{L_0}\right) = Z_V(g_0^2) + C_1\cdot\left(\frac{a}{L_0}\right)^2 
								+ C_2\cdot(a\LambdaQCD)\left(\frac{a}{L_0}\right)
								+ C_3\cdot(a\LambdaQCD)^2 + \;...
	\label{eq:ZVscaling}
\end{equation}
were the dots stand for higher order terms in the lattice spacing. The $(a/L_0)^2$ term is the dominant one thanks to the O($a$) improvement and to the fact that $a\LambdaQCD$ is a small factor in our finite temperature setup. The $(a\LambdaQCD)^2$ term is part of the definition of $Z_V$, and it vanishes quadratically in the lattice spacing when a renormalized correlator involving the flavor-singlet vector current is extrapolated to the continuum limit.

\begin{figure}
	\centering
	\begin{subfigure}{.5\textwidth}
		\centering
		\includegraphics[scale=0.45, trim=0cm 0.9cm 0cm 0cm clip]{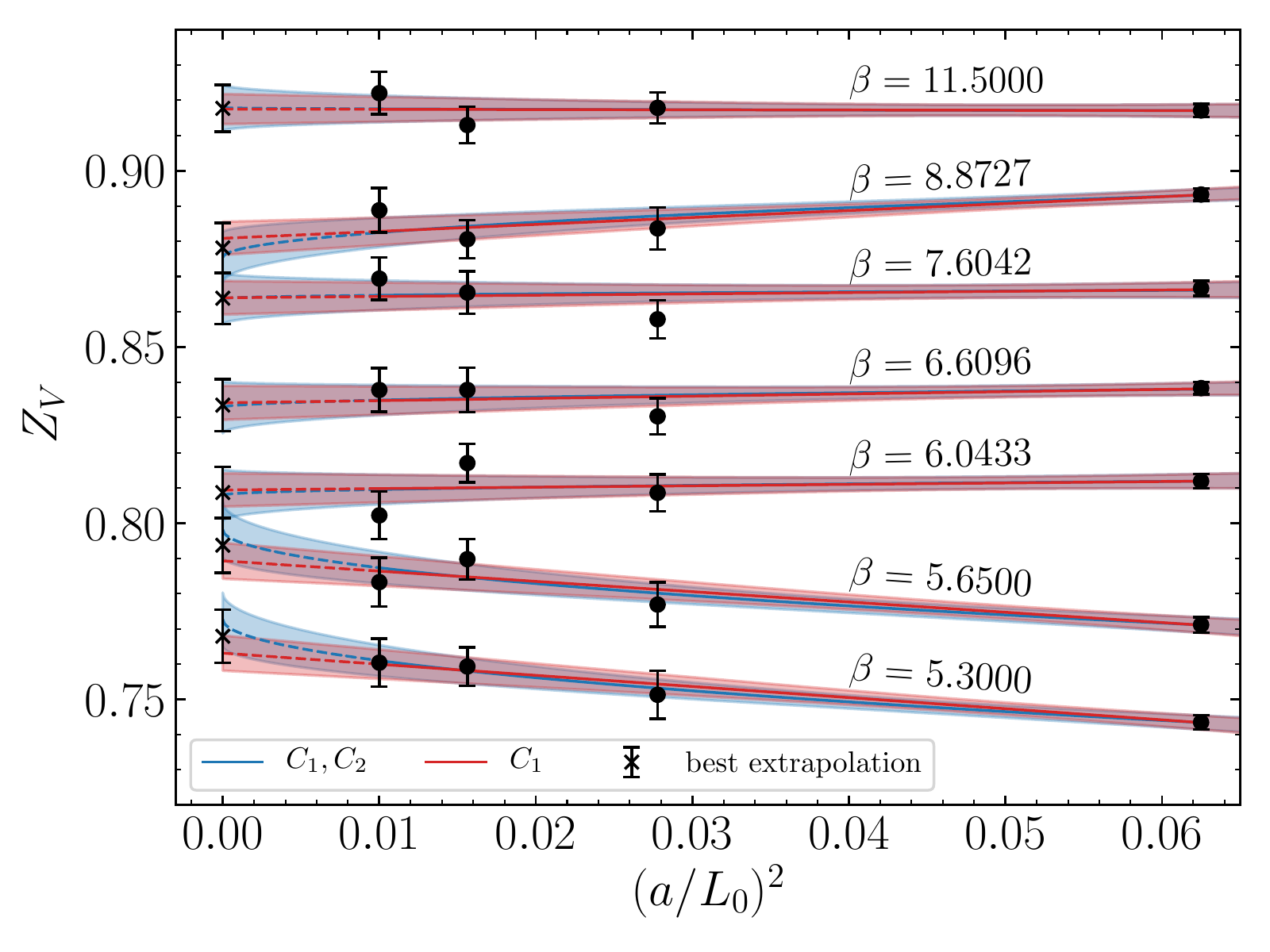}
	\end{subfigure}%
	\begin{subfigure}{.5\textwidth}
		\centering
		\includegraphics[scale=0.45, trim=0cm 0.9cm 0cm 0cm clip]{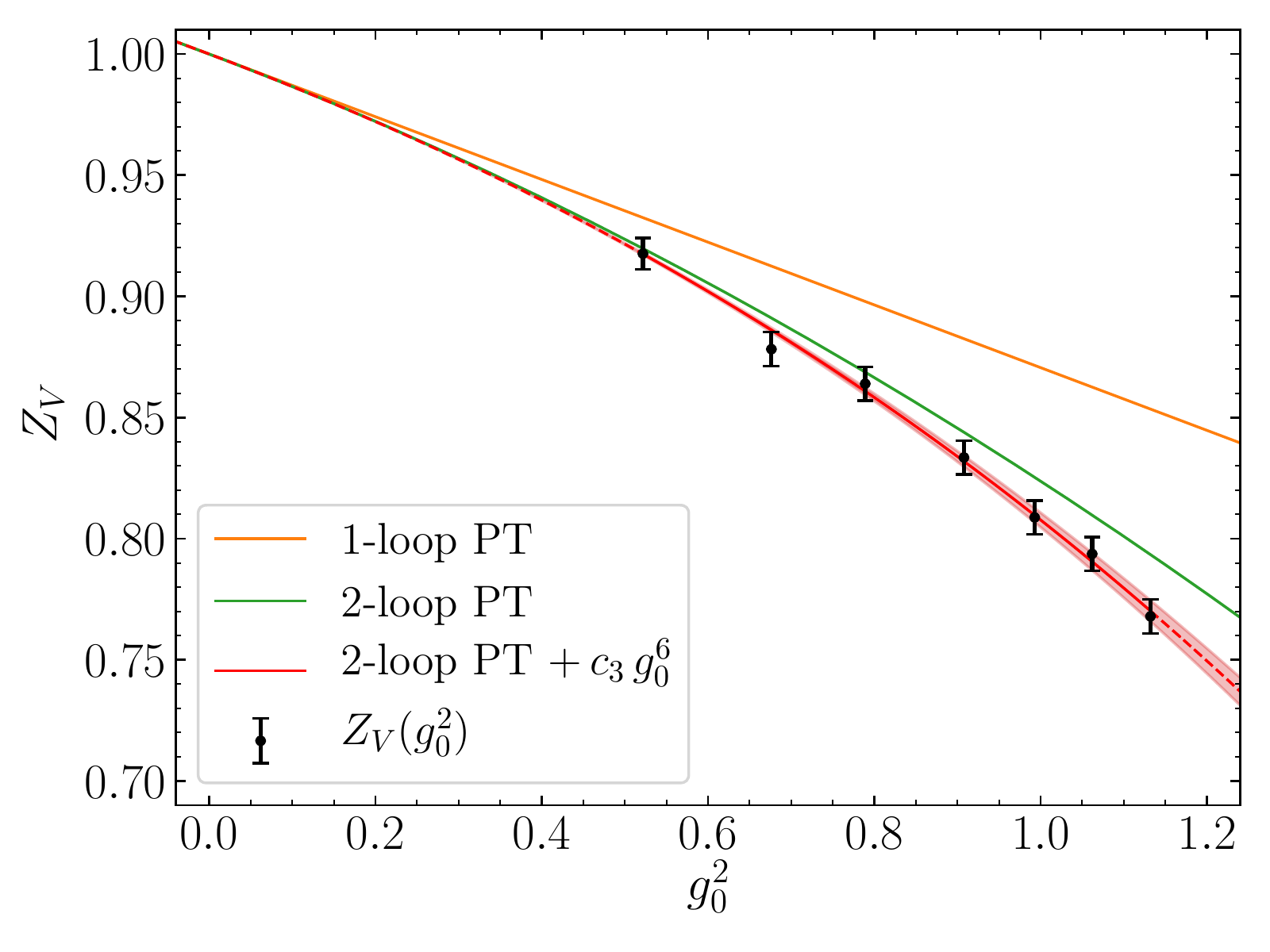}
	\end{subfigure}
	\caption{Left: extrapolations of numerical data in the limit $a/L_0\to0$. Right: comparison of extrapolated numbers with perturbation theory up to 2 loops.}
	\label{fig:ZV}
\end{figure}

The left plot in Figure \ref{fig:ZV} shows the numerical values we obtained from simulations of lattice QCD with spatial extension $L/a=96$ and $\Nf=3$ flavors of massless O($a$)-improved Wilson fermions.
We performed two extrapolations to the $a/L_0\to0$ limit, considering first both the $C_1,C_2$ terms of equation \eqref{eq:ZVscaling}, and then the $C_1$ term only. The extrapolated values are compatible, and this confirms that the residual linear dependence in $a/L_0$ in equation \eqref{eq:ZVscaling} is negligible for all the values of $\beta=6/g_0^2$ we considered. To be the most conservative, we average the two extrapolations at each $\beta$, and we take the largest error as the final uncertainty. The errors we obtain are less than $1\%$ and they are fully dominated by statistics.

The plot on the right in Figure \ref{fig:ZV} compares the numerical values of $Z_V$ with the perturbative prediction for the renormalization constant up to 2 loops \cite{Skouroupathis:2008mf}. The final result of this study is the polynomial interpolation
\begin{equation}
	Z_V^{\rm fit}(g_0^2) = 1 -0.129 g_0^2 -0.047 g_0^4 + c_3 \, g_0^6
	\label{eq:ZVinterp}
\end{equation}
where the coefficients up to $g_0^4$ come from perturbation theory (after checking that they were compatible with the fit) while $c_3=-0.016(3)$ comes from the numerical data and it is responsible of the mild bending of the non-perturbative points with respect to 2-loop perturbation theory at the larger values of bare coupling.

\section{Renormalization of the QCD energy-momentum tensor}
\label{sec:Z_Tmunu}
In the continuum the EM tensor of QCD is the quantity associated to the Poincaré symmetry and scale invariance of the theory. Because of the breaking of the translation and rotation symmetries by the lattice regularization, the 9-dimensional symmetric component of the continuum EM tensor splits in two bare lattice operators transforming in the triplet and sextet representations of the hypercubic group SW$_4$ \cite{Caracciolo:1989pt}. Each representation further splits in two contributions, one from the gluons and one from the fermions, and since there are no other lattice operators with the same symmetries and mass dimension $\leq 4$ we end up with four renormalization constants:
\begin{equation}
	T^{R,\{i\}}_{\mu\nu} = Z_G^{\{i\}} \; T^{G,\{i\}}_{\mu\nu} + Z_F^{\{i\}} \; T^{F,\{i\}}_{\mu\nu}, \mathsp i=3,6
	\label{eq:Tmunu}
\end{equation}
These renormalization constants are functions of the bare coupling $g_0$ only and they approach 1 as $g_0\to0$ (that is, in the continuum limit). We constrain their values by imposing the renormalized lattice EM tensor \eqref{eq:Tmunu} to satisfy a set of Ward Identities holding for the target operator in the continuum, up to cutoff effects.

Concerning the renormalization constants of the sextet representation, we consider the continuum relation \cite{DallaBrida:2020gux}
\begin{equation}
	\corr{T_{0k}}_\vxithzero = -\frac{\partial}{\partial \xi_k} f(L_0, L, \vxi, \theta_0)
	\label{eq:Tcond1}
\end{equation}
between the EM tensor and the derivative with respect to the shift of the free energy density of the thermal system
\begin{equation}
	f(L_0, L, \vxi, \theta_0) = -\frac{1}{L_0L^3} \ln Z\,, \mathsp 
	Z\left[L_0, L, \vxi, \theta_0\right] = \int DU D\psibar D\psi\; e^{-S\left[U,\psibar,\psi\right]}
	\label{eq:free_energy}
\end{equation}
and we impose \eqref{eq:Tcond1} on the renormalized lattice sextet EM tensor up to discretization errors. Note that this relation is non-trivial in presence of shifted boundary conditions only. Moreover, we evaluate the very same relation at two twist phases $\theta_0^A$ and $\theta_0^B$ for the fermions on the temporal boundary, so that we obtain the two constraints needed for solving the mixing  between the gluonic and fermionic components:
\begin{equation}
	\begin{cases}
		\Zgs \corr{\Tgs[0k]}_{\vxi,\theta_0^A} + \Zfs \corr{\Tfs[0k]}_{\vxi,\theta_0^A} = 
		   - \dfrac{\Delta f(L_0, L,\vxi,\theta_0^A)}{\Delta\xi_k} + \text{O}(a^2)&\\ \\
		\Zgs \corr{\Tgs[0k]}_{\vxi,\theta_0^B} + \Zfs \corr{\Tfs[0k]}_{\vxi,\theta_0^B} = 
		   - \dfrac{\Delta f(L_0, L,\vxi,\theta_0^B)}{\Delta\xi_k} +\text{O}(a^2)&\\
	\end{cases}
	\label{eq:Tsys}
\end{equation}
The renormalization constants of the triplet EM tensor can be computed similarly thanks to the Ward Identity \cite{DallaBrida:2020gux}
\begin{equation}
	    \corr{T^{R,\{6\}}_{0k}}_{\vxi,\theta_0} = \xi_k\corr{T^{R,\{3\}}_{0j}}_{\vxi,\theta_0} 
	    \mathsp (j\neq k, \xi_j=0)
	\label{eq:Tcond2}
\end{equation}
Again, this Ward Identity is non-trivial in a shifted frame only.
The unknown variables of the linear system in equation \eqref{eq:Tsys} are the renormalization constants $Z_G^{\{6\}},\;Z_F^{\{6\}}$. The one-point functions of the bare lattice EM tensor, as well as the discrete derivatives of the free energy density, must be determined from lattice simulations. In the following we give some results for the former ones, while the derivatives of the free energy are work in progress.

Our renormalization strategy relies on the possibility of evaluating equation \eqref{eq:Tcond1} at two different values of $\theta_0$. It is therefore important to check that we can resolve with sufficient precision the relevant one-point functions of the EM tensor at two imaginary chemical potentials, and Table \ref{tab:Tmunudata} tells us that this is the case. In the latter we collect some preliminary numerical results for the lowest and highest values of $\beta$ we are considering. We simulate $N_f=3$ massless flavors of O($a$)-improved Wilson fermions, on a lattice of spatial extension $L/a=288$ and temporal one $L_0/a=4,6$ but we are planning simulations at $L_0/a=8,10$ too, in order to perform the $a/L_0\to 0$ extrapolation. We see that the one-point functions have a relative error of about $1\%$, and they are remarkably different at the two values $\theta_0^A=0$, $\theta_0^B=\pi/6$ even for the gluonic components, which depend on the chemical potential as a 1-loop effect.
\begin{table}
	\centering
	\begin{tabular}{cccccccc}
	\hline
	$\theta_0$ & $\beta$ & $L_0/a$ & $N_{\rm{traj}}$ & $\corr{\Tgs}/T^4$ & $\corr{\Tfs}/T^4$ 
	                               & $\corr{\Tgt}/T^4$ & $\corr{\Tft}/T^4$ \\
	\hline
	\multirow{4}{*}{$0$}
	& \multirow{2}{*}{6.0433}  & 4 & 50 & -2.325(8)  & -6.335(3) & -2.760(11) & -7.081(5)  \\
	&	 					   & 6 & 100 & -2.314(19) & -5.772(8) & -2.63(3)   & -6.303(15)
	\vspace{0.2cm}\\
	& \multirow{2}{*}{8.8727}  & 4 & 50 & -2.842(8) &  -6.982(3) & -3.223(12) & -7.692(4)  \\
	&	 					   & 6 & 100 & -2.822(24) & -6.343(9) & -3.13(4)   & -6.846(11) \\
	\hline
	\multirow{4}{*}{$\pi/6$}
	& \multirow{2}{*}{6.0433} & 4 & 50 & -2.247(5)  & -5.5939(21) & -2.642(7) & -6.2617(29)  \\
	&                         & 6 & 100 & -2.198(19) & -5.115(9)   & -2.59(4)  & -5.587(15)
	\vspace{0.2cm}\\
	& \multirow{2}{*}{8.8727} & 4 & 50 & -2.784(7)  & -6.1605(29) & -3.164(12) & -6.783(4)  \\
	&                         & 6 & 100 & -2.753(26) & -5.615(7)   & -3.00(4)   & -6.077(13) \\
	\hline
	\end{tabular}
	\caption{Numerical values of the relevant one-point functions for the renormalization of the energy-momentum tensor.}
	\label{tab:Tmunudata}
\end{table}

\section{Conclusions}
Thermal QCD in a moving frame and with an imaginary chemical potential revealed to be an effective framework for the non-perturbative renormalization of the flavor-singlet local vector current \cite{Bresciani:2022lqc}. Taking advantage of all the properties of this new renormalization scheme we are now working on the non-perturbative renormalization of the energy-momentum tensor of QCD.
This result will allow the study from first principles of many fundamental properties of QCD at finite temperature, such as the Equation of State of QCD or the transport coefficients in a Quark-Gluon Plasma \cite{Meyer:2011gj}. The Equation of State has been non-perturbatively determined in the past with staggered quarks \cite{HotQCD:2014kol,Bazavov:2017dsy} through the numerical computation of the trace anomaly $\varepsilon(T)-3p(T)$, whose integral in the temperature gives access to the pressure density $p(T)$. Then the energy density $\varepsilon(T)$ and the entropy density $s(T)$ come from standard thermodynamic relations. However this approach, known as integral method, becomes numerically very challenging at high temperatures since the trace anomaly, having a quartic divergence in the continuum limit, must be subtracted for instance by its value at zero temperature. For this reason the available numerical results reach temperatures of the order of 1 GeV at most. This calls for a change of strategy, also because perturbation theory is known to be unreliable at higher temperatures at least in the SU$(3)$ Yang-Mills theory \cite{Giusti:2016iqr}.

We are planning to determine the QCD Equation of State from first principles with O($a$)-improved Wilson fermions using the equation
\begin{equation}
	\frac{s}{T^3} = - L_0^4\, \frac{(1+\vxi^2)^3}{\xi_k}\corr{T^R_{0k}}\; , 
				\mathsp T = \frac{1}{L_0\sqrt{1+\vxi^2}}
	\label{eq:entropy}
\end{equation}
which directly relates the entropy density $s$ of the Quark-Gluon Plasma at temperature $T$ in a moving reference frame with Euclidean speed $\vxi$ with a one-point function of the spacetime components of the renormalized energy-momentum tensor \cite{DallaBrida:2020gux}. Crucially these components do not need any zero point subtraction, and this opens the way to the determination of the QCD Equation of State in the unexplored temperature range from 1 GeV up to the electroweak scale.


\begin{thebibliography}{1}

\bibitem{Luscher:1992an}
M.~Luscher, R.~Narayanan, P.~Weisz and U.~Wolff,
Nucl. Phys. B \textbf{384} (1992), 168-228
doi:10.1016/0550-3213(92)90466-O
[arXiv:hep-lat/9207009 [hep-lat]].

\bibitem{Martinelli:1994ty}
G.~Martinelli, C.~Pittori, C.~T.~Sachrajda, M.~Testa and A.~Vladikas,
Nucl. Phys. B \textbf{445} (1995), 81-108
doi:10.1016/0550-3213(95)00126-D
[arXiv:hep-lat/9411010 [hep-lat]].

\bibitem{Luscher:2010iy}
M.~L\"uscher,
JHEP \textbf{08} (2010), 071
[erratum: JHEP \textbf{03} (2014), 092]
doi:10.1007/JHEP08(2010)071
[arXiv:1006.4518 [hep-lat]].

\bibitem{Giusti:2010bb}
L.~Giusti and H.~B.~Meyer,
Phys. Rev. Lett. \textbf{106} (2011), 131601
doi:10.1103/PhysRevLett.106.131601
[arXiv:1011.2727 [hep-lat]].

\bibitem{DallaBrida:2020gux}
M.~Dalla Brida, L.~Giusti and M.~Pepe,
JHEP \textbf{04} (2020), 043
doi:10.1007/JHEP04(2020)043
[arXiv:2002.06897 [hep-lat]].

\bibitem{Giusti:2011kt}
L.~Giusti and H.~B.~Meyer,
JHEP \textbf{11} (2011), 087
doi:10.1007/JHEP11(2011)087
[arXiv:1110.3136 [hep-lat]].

\bibitem{Giusti:2012yj}
L.~Giusti and H.~B.~Meyer,
JHEP \textbf{01} (2013), 140
doi:10.1007/JHEP01(2013)140
[arXiv:1211.6669 [hep-lat]].

\bibitem{Giusti:2015daa}
L.~Giusti and M.~Pepe,
Phys. Rev. D \textbf{91} (2015), 114504
doi:10.1103/PhysRevD.91.114504
[arXiv:1503.07042 [hep-lat]].

\bibitem{Giusti:2016iqr}
L.~Giusti and M.~Pepe,
Phys. Lett. B \textbf{769} (2017), 385-390
doi:10.1016/j.physletb.2017.04.001
[arXiv:1612.00265 [hep-lat]].

\bibitem{Bresciani:2022lqc}
M.~Bresciani, M.~D.~Brida, L.~Giusti, M.~Pepe and F.~Rapuano,
Phys. Lett. B \textbf{835} (2022), 137579
doi:10.1016/j.physletb.2022.137579
[arXiv:2203.14754 [hep-lat]].

\bibitem{Hatton:2020vzp}
D.~Hatton \textit{et al.} [HPQCD],
Phys. Rev. D \textbf{102} (2020) no.9, 094509
doi:10.1103/PhysRevD.102.094509
[arXiv:2008.02024 [hep-lat]].

\bibitem{Roberge:1986mm}
A.~Roberge and N.~Weiss,
Nucl. Phys. B \textbf{275} (1986), 734-745
doi:10.1016/0550-3213(86)90582-1

\bibitem{Bhattacharya:2005rb}
T.~Bhattacharya, R.~Gupta, W.~Lee, S.~R.~Sharpe and J.~M.~S.~Wu,
Phys. Rev. D \textbf{73} (2006), 034504
doi:10.1103/PhysRevD.73.034504
[arXiv:hep-lat/0511014 [hep-lat]].

\bibitem{Skouroupathis:2008mf}
A.~Skouroupathis and H.~Panagopoulos,
Phys. Rev. D \textbf{79} (2009), 094508
doi:10.1103/PhysRevD.79.094508
[arXiv:0811.4264 [hep-lat]].

\bibitem{Luscher:1996vw}
M.~Luscher and P.~Weisz,
Nucl. Phys. B \textbf{479} (1996), 429-458
doi:10.1016/0550-3213(96)00448-8
[arXiv:hep-lat/9606016 [hep-lat]].

\bibitem{Caracciolo:1989pt}
S.~Caracciolo, G.~Curci, P.~Menotti and A.~Pelissetto,
Annals Phys. \textbf{197} (1990), 119
doi:10.1016/0003-4916(90)90203-Z

\bibitem{Meyer:2011gj}
H.~B.~Meyer,
Eur. Phys. J. A \textbf{47} (2011), 86
doi:10.1140/epja/i2011-11086-3
[arXiv:1104.3708 [hep-lat]].

\bibitem{HotQCD:2014kol}
A.~Bazavov \textit{et al.} [HotQCD],
Phys. Rev. D \textbf{90} (2014), 094503
doi:10.1103/PhysRevD.90.094503
[arXiv:1407.6387 [hep-lat]].

\bibitem{Bazavov:2017dsy}
A.~Bazavov, P.~Petreczky and J.~H.~Weber,
Phys. Rev. D \textbf{97} (2018) no.1, 014510
doi:10.1103/PhysRevD.97.014510
[arXiv:1710.05024 [hep-lat]].



\end{thebibliography}
\end{document}